\documentclass{PoS}

\title{Single and double diffractive prompt photon production at the LHC}

\ShortTitle{Single and double diffractive prompt photon production at the LHC}

\author{\speaker{Cristiano Brenner Mariotto}\thanks{This work was partially financed by the Brazilian funding agencies FAPERGS, CNPq and CAPES.}         \\
        Universidade Federal do Rio Grande (FURG), Rio Grande, Brazil\\
        E-mail: \email{cristianomariotto@furg.br}}

\author{V. P. Goncalves\\
Universidade Federal de Pelotas (UFPEL), Pelotas, Brazil \\
        E-mail: \email{barros@ufpel.edu.br}}

\abstract{In this contribution we study for the first time the prompt photon production in single and double diffractive processes considering the resolved Pomeron model. We estimate the rapidity and transverse momentum dependence of the cross section for the production of two photons and of a photon and a jet. We show that these processes are sensitive to the pomeron structure. In contrast with the dijet and heavy quark production, which are dominated by gluon-gluon interactions, in prompt photon production, Compton like processes are dominant for single photon plus jet events. This gives a unique oportunity to constrain the quark distribution in the Pomeron. The results are obtained for the LHC energy.}

\FullConference{XXI International Workshop on Deep-Inelastic Scattering and Related Subjects\\
		 22-26 April, 2013\\
		 Marseilles, France}
\def\pom{{I\!\!P}}

\begin{document}

\section{Introduction}
A long-standing puzzle in the particle physics is the nature of the Pomeron ($\pom$). This object, with the vacuum quantum numbers, was introduced phenomenologically in the Regge theory as a simple moving pole in the complex angular momentum plane, to describe the high-energy behavior of the total and elastic cross-sections of the hadronic reactions \cite{collins}. Due to its zero color charge the Pomeron is associated with diffractive events, characterized by the presence of large rapidity gaps in the hadronic final state. The diffractive processes have attracted much attention as a way of amplifying the physics programme at hadronic colliders, including searching for New Physics (For a recent review see, e.g. Ref. \cite{forshaw}). The investigation of these reactions at high energies gives important information about the structure of hadrons and their interaction mechanisms. 

The diffractive processes can be classified as inclusive or exclusive events (See e.g. \cite{forshaw}).
In exclusive events, empty regions in pseudo-rapidity, called rapidity gaps, separate the intact very forward hadron from the central massive object. Exclusivity means that nothing else is produced except the leading hadrons and the central object. 
The inclusive processes also exhibit rapidity gaps. However, they contain soft particles accompanying the production of a hard diffractive object, with the rapidity gaps becoming, in general, smaller than in the exclusive case. Moreover, 
the inclusive  processes can also be classified as single or double diffractive, which is directly associated to the presence of one or two rapidity gaps in the final state, respectively.

The diffractive physics has been tested in hadron-hadron collisions considering  distinct final states like dijets, dileptons, heavy quarks, quarkonium + photon, and different theoretical approaches. One of these approaches is the Resolved Pomeron Model, which assumes the validity of the diffractive factorization formalism and that Pomeron has a partonic structure. The basic idea is that the hard scattering resolves the quark and gluon content in the Pomeron \cite{IS} and it can be obtained analysing the experimental data from diffractive deep inelastic scattering (DDIS) at HERA, providing us with the diffractive distributions of singlet quarks and gluons in the Pomeron \cite{H1diff}. However, other approaches based on very distinct assumptions,  for example the BFKL Pomeron, are also able to describe the current scarce experimental data. Consequently, the present scenario for diffractive processes is unclear, motivating the study of alternative processes which allow to constrain the correct description of the Pomeron.

In this contribution we study the prompt photon production as a complementary test of diffractive processes and the pomeron structure. We stress that this process has still not been calculated in the literature, and this is the first approach to that. 
 The production of prompt photons in nondiffractive processes in hadron-hadron collisions is considered usually as a  probe of the proton's gluon distribution due to dominance of the LO Compton-like subprocess $q g \rightarrow \gamma q$ (See, e.g., \cite{crisvic_foton}). Our goal is to extend there previous analysis for diffractive processes considering parton - Pomeron and Pomeron - Pomeron interactions and predict the pseudorapidity and transverse momentum dependences, as well as the total cross section, for the Photon + jet and double photon production (For a more detailed study see Ref. \cite{Mariotto:2013kca}).

\section{Single and Double Diffractive prompt photon production}

At leading order the prompt photon production is determined by 
 the Compton process $qg \rightarrow q \gamma$, which is dominant at high energies, the annihilation process  $q \bar q  \rightarrow g \gamma$ and the processes  $q \bar q  \rightarrow \gamma \gamma$ (pure EM), $g g  \rightarrow \gamma \gamma$ and $g g  \rightarrow g \gamma$, which are subdominant. In order to estimate the hadronic cross sections we have to convolute the cross sections for these partonic subprocesses with the inclusive and/or diffractive parton distribution functions. In the Resolved Pomeron Model, the diffractive parton distributions in the proton are defined as a convolution of the Pomeron flux emitted from the proton, $f_{I\!\!P}(x_{I\!\!P})= \int_{t_{min}}^{t_{max}} dt f_{{I\!\!P}}(x_{{I\!\!P}}, t)$, and the parton distributions in the Pomeron, $g_{I\!\!P}(\beta, \mu^2)$, $q_{I\!\!P}(\beta, \mu^2)$, where $\beta$ is the momentum fraction carried by the partons inside the Pomeron.  
The diffractive quark and gluon distributions are given by
\begin{eqnarray}
{ q^D(x,\mu^2)}=\int dx_{I\!\!P}d\beta \delta (x-x_{I\!\!P}\beta)f_{I\!\!P}(x_{I\!\!P})q_{I\!\!P}(\beta, \mu^2)={ \int_x^1 \frac{dx_{I\!\!P}}{x_{I\!\!P}} f_{I\!\!P}(x_{I\!\!P}) q_{I\!\!P}\left(\frac{x}{x_{I\!\!P}}, \mu^2\right)} \\
{ g^D(x,\mu^2)}=\int dx_{I\!\!P}d\beta \delta (x-x_{I\!\!P}\beta)f_{I\!\!P}(x_{I\!\!P})g_{I\!\!P}(\beta, \mu^2)={ \int_x^1 \frac{dx_{I\!\!P}}{x_{I\!\!P}} f_{I\!\!P}(x_{I\!\!P}) g_{I\!\!P}\left(\frac{x}{x_{I\!\!P}}, \mu^2\right)}
\end{eqnarray}

\begin{figure}[t]
\begin{center}
\scalebox{0.30}{\includegraphics{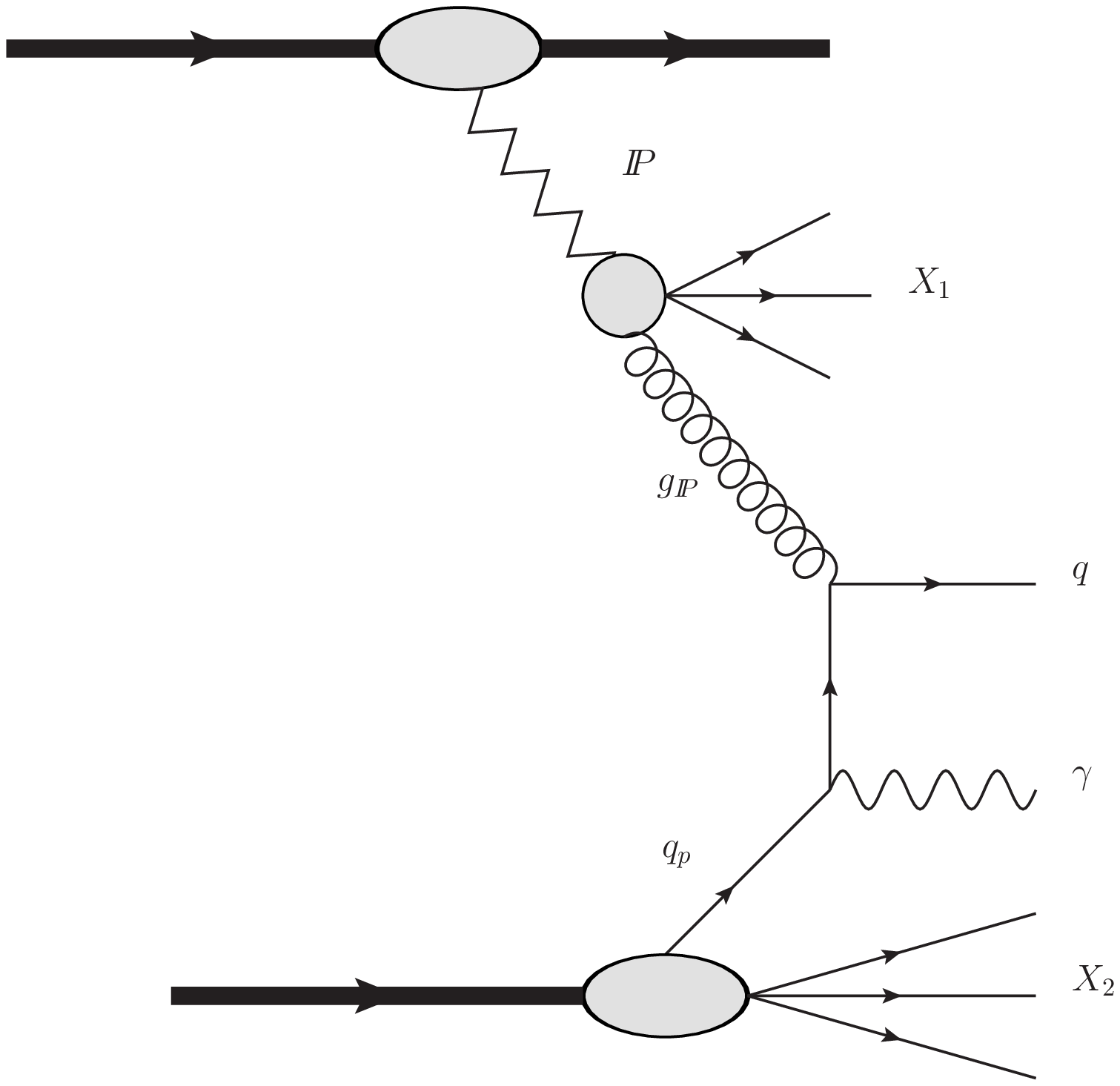}}
\scalebox{0.25}{\includegraphics{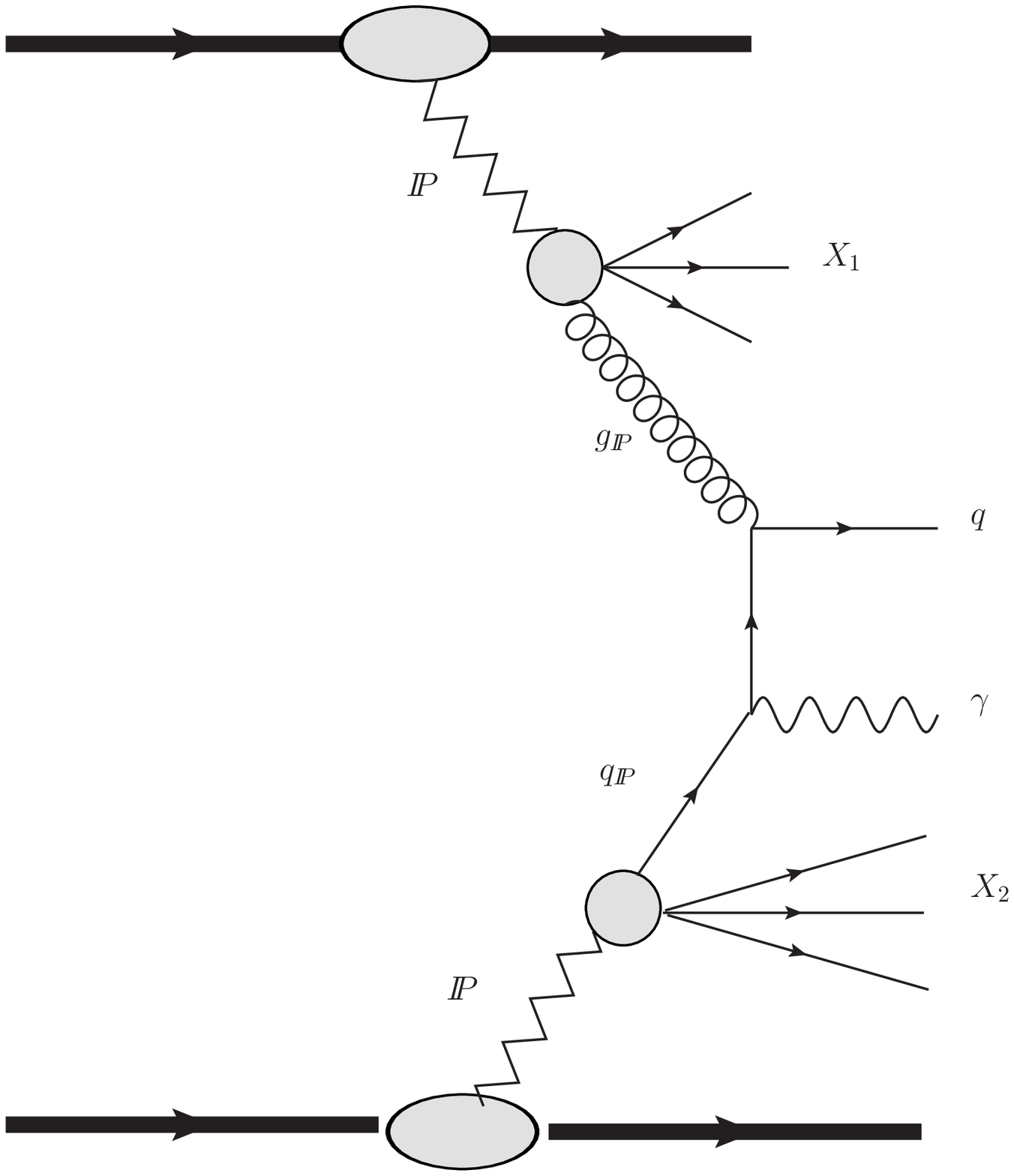}}
\caption{Photon + jet production in single (left panel) and double (right panel) diffractive processes.}
\label{single}
\end{center}
\end{figure}

\begin{figure}[t]
\begin{center}
\scalebox{0.30}{\includegraphics{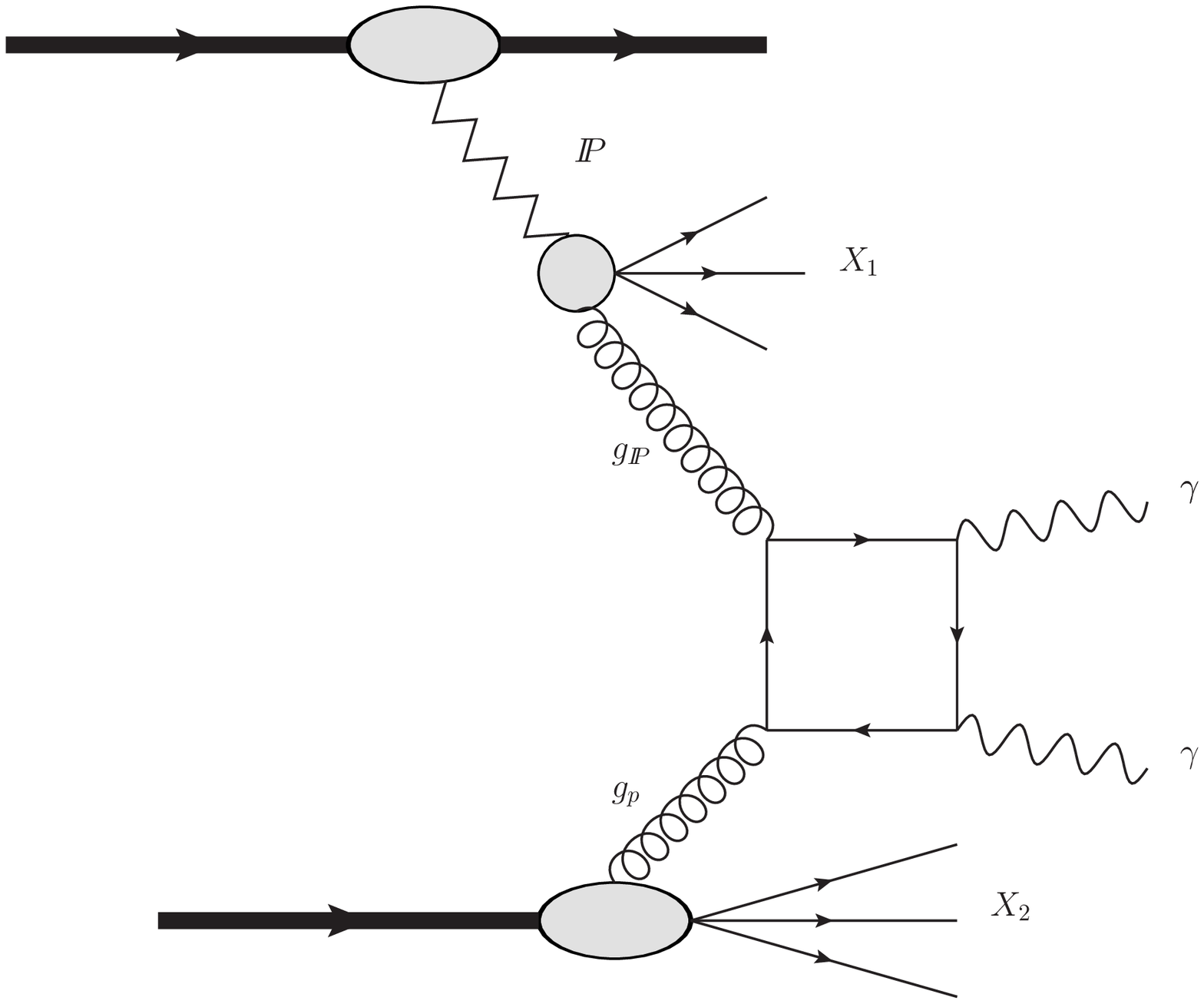}}
\scalebox{0.25}{\includegraphics{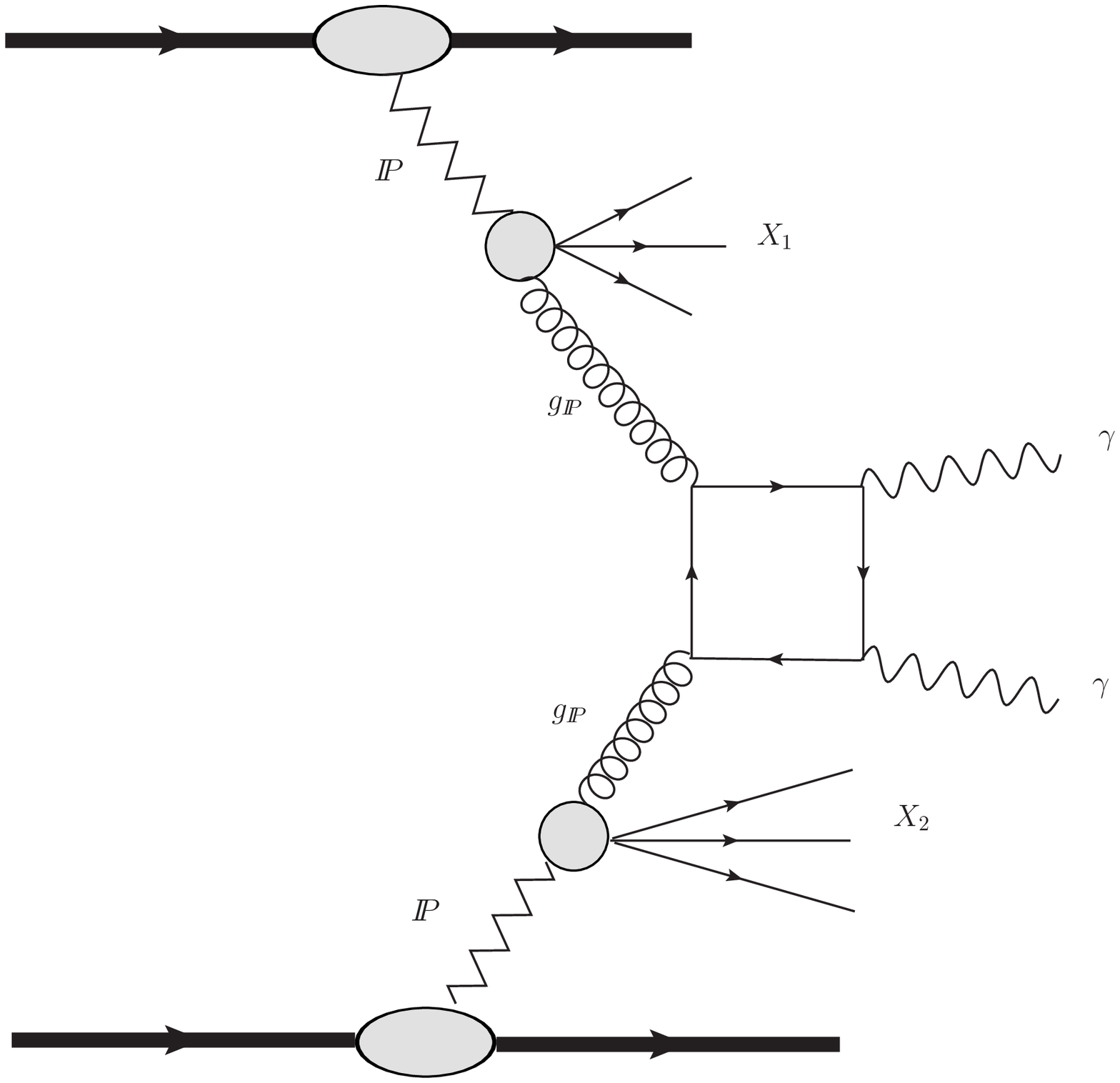}}
\caption{Double photon production in single (left panel) and double (right panel) diffractive processes.}
\label{figsdddouble}
\end{center}
\end{figure}

The Photon + jet production in single and double diffractive processes is described by the diagrams like those presented in Fig. 
\ref{single}, where we have the  production of a photon plus a jet or an unobserved photon ($\gamma X$). In single prompt photon diffractive production the Pomeron might be emitted from one of the two protons and one should include both $pI\!\!P$ and $I\!\!Pp$ interactions. The corresponding cross section may be written as
 \begin{eqnarray}
\frac{d\sigma }{dydp_T^2}= \sum_{abcd} \int_{x_{a\, min}}^1 dx_a
{f^D_a(x_a,Q^2)}f_b(x_b,Q^2)\frac{x_ax_b}{2x_a-x_Te^y}\frac{d\hat{\sigma}}{d\hat{t}}(ab\rightarrow
cd)
\end{eqnarray}
where in fact one considers ${f^D_{a}}$ $f_{b/p}$ $+$ $f_{a/p}$ ${f^D_{b}}$ contributions, 
$x_{a\, min}=\frac{x_Te^{y}}{2-x_Te^{-y} }$,
$x_b=\frac{x_ax_Te^{-y}}{2x_a-x_Te^y}$, 
$x_T=2p_T/\sqrt{s}$ and $\frac{d\hat{\sigma}}{d\hat{t}}$ are the LO partonic cross sections.
In the case of double diffractive prompt photon production (also called central diffractive)  one has the ${I\!\!P} {I\!\!P} $ interactions (${f^D_{a}}{f^D_{b}}$ in the formula above) and one looks for events with two rapidity gaps and a $\gamma +$jet at the central region. 
Moreover, we consider the double photon production in single and double diffractive processes, which are described by the diagrams like those shown in Fig. \ref{figsdddouble}, taking into account the $g g  \rightarrow \gamma \gamma$ and  $q \bar q  \rightarrow \gamma \gamma$ subprocesses.

\section{Results}
\begin{figure}[t]
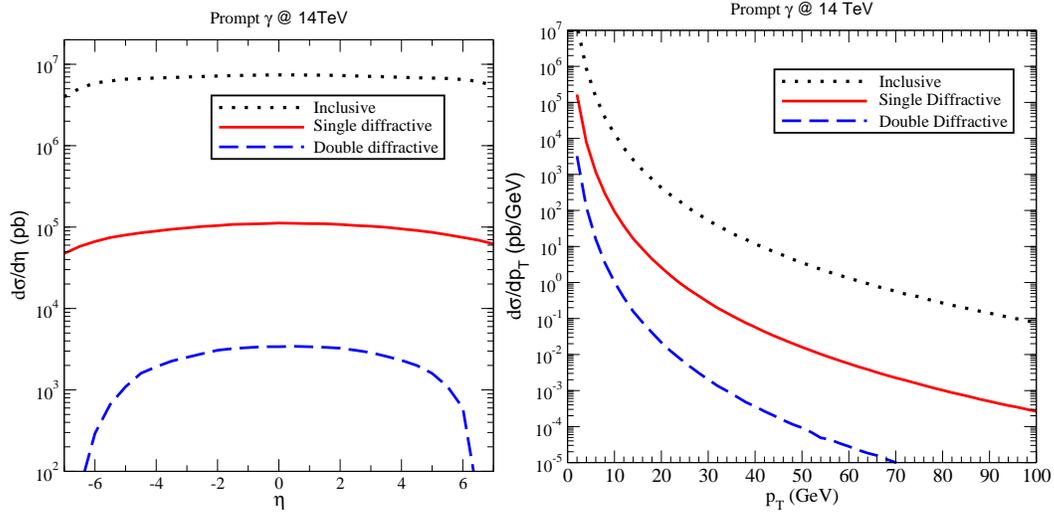

\begin{center}
\scalebox{0.36}{\includegraphics{promptgamma_eta.eps}}
\scalebox{0.36}{\includegraphics{promptgamma_pt.eps}}
\caption{Pseudorapidity (left panel) and transverse momentum (right panel) distributions for the prompt photon production at $\sqrt{s}= 14$ TeV.}
\label{figsdddsigle}
\end{center}
\end{figure}

\begin{figure}[t]
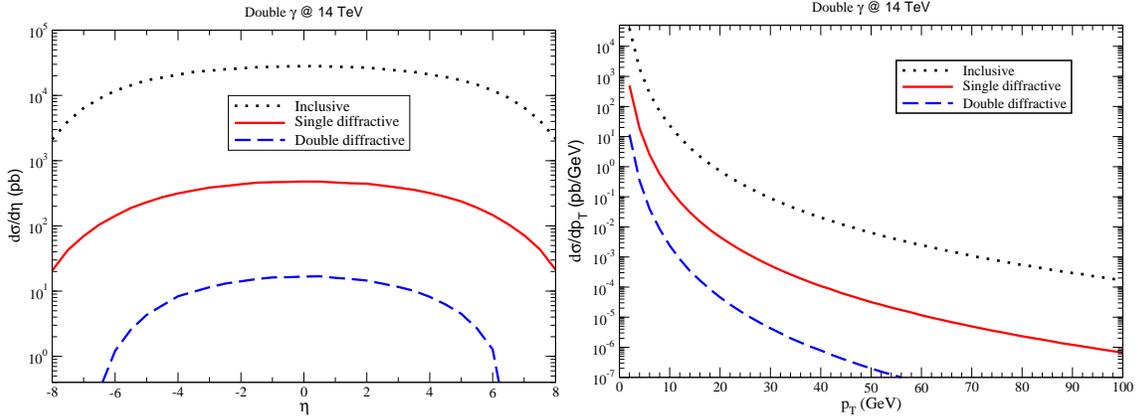

\begin{center}
\scalebox{0.31}{\includegraphics{doublegamma_etav2.eps}}
\scalebox{0.31}{\includegraphics{doublegamma_ptv2.eps}}
\caption{Pseudorapidity (left panel) and transverse momentum (right panel) distributions for the double photon production at $\sqrt{s}= 14$ TeV.}\label{figsddddouble}
\end{center}
\end{figure}

In what follows we present our results for total cross sections, the pseudorapidity $\eta$ and transverse momentum $p_T$ distributions at the LHC energy of $\sqrt{s} = 14$ TeV. For the sake of comparison, we also show the inclusive (non-diffractive) LO contribution.
Following previous studies for single and double diffractive production \cite{mag_hqdif,MMM1,schurek} we also include in our calculations a gap survival factor given by 0.05 (0.02) for single (double) diffraction. 
In Tab. \ref{tab1} we present our predictions for the total cross sections.
One observes a reduction of two orders of magnitude
when going from the inclusive to single diffractive, and from single to double diffractive.
This behaviour is also observed in the differential distributions presented in Figs. \ref{figsdddsigle} and \ref{figsddddouble}. For prompt photon production, the cross sections are sizable and could be measured at the LHC. For double photon production, one has much cleaner processes than for prompt photon production because of the absence of a produced jet. On the other hand, the cross sections are much smaller (by three orders of magnitude), since only the subdominant subprocesses contribute. Anyway, these four extra channels considered in this work might be usefull to study the Pomeron structure and interactions.

\begin{table}[h]
\begin{center}
 \vspace{0.5cm}
\begin{tabular}{|c|c|c|c|}
\hline
\hline
Final state & Inclusive & Single diffractive & Double diffractive  \\
\hline
\hline
 $\gamma $ jet & $1.02\times 10^8$ pb & 1.37$\times 10^6$ pb &  2.95$\times 10^4$ pb \\
\hline
 $\gamma \gamma$ & $2.98\times 10^5$ pb & 4617.0 pb  & 128.0  pb  \\
  \hline
\hline
\end{tabular}
\caption{Total cross sections for photon+jet and double photon production.}
\label{tab1}
\end{center}
\end{table}

As a summary, in this contribution we have presented for the first time predictions for diffractive production of photons in $pp$ collisions at the LHC. Our results indicate that the experimental analysis is feasible and it would help in constraining the underlying model for the Pomeron and the diffractive parton distributions.

\end{document}